\documentclass[manuscript]{aastex6}
\newcommand{\wind}{\textit{Wind}}
\newcommand{\sdo}{\textit{SDO}}

\newcommand{\stereo}{\textit{STEREO}}
\newcommand{\sta}{\textit{STEREO A}}
\newcommand{\stb}{\textit{STEREO B}}

\newcommand{\soho}{\textit{SOHO}}

\newcommand{\soholasco}{\textit{SOHO${/}$}LASCO}
\newcommand{\insitu}{\textit{in situ}}

\usepackage{amsmath}
\shorttitle{Observaions of the Rotation and non-radial motion of MFR}
\shortauthors{Liu et al.}
\begin{document}
\title{Multi-Spacecraft Observations of the Rotation and Non-Radial Motion of a CME  Flux Rope causing an intense geomagnetic storm}
\author{Yi A. Liu\altaffilmark{1,2}, Ying D. Liu\altaffilmark{1,2},
        Huidong Hu\altaffilmark{1}, Rui Wang\altaffilmark{1,3}, and Xiaowei Zhao\altaffilmark{1,2}}
\altaffiltext{1}{State Key Laboratory of Space Weather,
        National Space Science Center,
        Chinese Academy of Sciences, Beijing 100190, China;
        \href{mailto:liuxying@swl.ac.cn}{liuxying@swl.ac.cn}}
\altaffiltext{2}{University of Chinese Academy of Sciences, Beijing 100049, China}
\altaffiltext{3}{W. W. Hansen Experimental Physics Laboratory, Stanford University, Stanford, CA, USA}

\begin{abstract}
  We present an investigation of the rotation and non-radial motion of a coronal mass ejection (CME) from  AR 12468 on 2015 December 16 using observations from \sdo, \soho, \sta{} and \wind. The EUV and HMI observations of the source region  show that the associated magnetic flux rope (MFR) axis pointed to the east before the eruption. We use a nonlinear fore-free field (NLFFF) extrapolation to determine the configuration of the coronal magnetic field and calculate the magnetic energy density distributions at different heights. The distribution of the magnetic energy density shows a strong gradient toward the northeast. The propagation direction of the CME from a Graduated Cylindrical Shell (GCS) modeling deviates from the radial direction of the source region by about 45{\degr} in longitude and about 30{\degr}  in latitude, which is consistent with the gradient of the magnetic energy distribution around the AR. The MFR axis determined by the GCS modeling points southward, which has  rotated counterclockwise by about 95{\degr} compared with the orientation of the MFR in the low corona.  The MFR reconstructed by a Grad-Shafranov (GS) method at 1 AU has almost the same orientation as the MFR from the GCS modeling, which indicates that the MFR rotation occurred in the low corona. It is the rotation of the MFR that caused the intense geomagnetic storm with the minimum D$_\mathrm{st}$ of --155 nT. These results suggest that the  coronal magnetic field surrounding the MFR plays a crucial role in the MFR rotation and propagation direction.
\end{abstract}

\section{Introduction}\label{intro}
Coronal mass ejections (CMEs) are large-scale eruptions of plasma and magnetic field from the Sun and  are a major driver of space weather effects. When propagating in the heliosphere, they are called interplanetary CMEs (ICMEs). 
A  previous  study suggests  that at least 40\%  of CMEs observed by \soholasco{} have a  clear magnetic flux rope (MFR) structure \citep{vourlidas2013}. In the low corona of the active region (AR), an MFR structure is manifested as a hot channel (HC), a high-temperature erupting structure seen in EUV observations \citep[e.g.,][]{cheng2013,cheng2014a,zhangj2012}. When a CME propagates from the Sun, it has a three-part structure in white-light observations: the white leading edge, core and dark cavity \citep{illing1985,forsyth2006}. The dark cavity is thought to be the MFR structure. The propagation direction of a CME MFR  will determine whether it can arrive at Earth, and the orientation  of a CME MFR  will determine its southward  magnetic field component at the Earth \citep[e.g.,][]{liu2010,liu2016,savani2015,kilpua2012,isavnin2014}. Therefore, understanding the propagation and rotation of a CME MFR in the inner heliosphere is of crucial importance for space weather forecasting.  

The propagation direction of a CME can  deviate from the radial direction of the CME source region. The magnetic force produced by the coronal magnetic field plays an important  role in CME propagation direction \citep{macqueen1986,kilpua2009}. \citet{shen2011} and \citet{gui2011}  use the gradient of coronal magnetic energy density to explain the deflection of a CME in the low corona and suggest that the deflection is  toward  the region of the lower magnetic energy. \citet{kay2013,kay2015} develop a Forecasting CMEs Altered Trajectory (ForeCAT) model to predict CME trajectory. Heliospheric current sheets (HCSs) and coronal holes (CHs), which correspond to minimum and maxmum  energy regions on the solar surface, respectively, are important factors that can deflect CMEs \citep{kilpua2009,gopalswamy2009,liewer2015}.  Recent studies have demonstrated that CMEs can be deflected not only by the nearby large-scale structure, such as HCSs and CHs, but also can be channelled by the strong magnetic field in the CME source region itself \citep{mostl2015,wang2015,huhd2017}. CMEs can also change their propagation direction in interplanetary space by interacting with fast solar wind streams and other CMEs \citep[e.g.,][]{liu2014nc,liu2016,lugaz2012,gopals2001,kataoka2015}. 

 CME MFRs have been observed to rotate frequently \citep[e.g.,][]{vourlidas2011,thompson2012,liu2008rot,liu2010}. \citet{liu2010} suggest that a CME can exhibit a significant rotation during its propagation in the inner heliosphere by comparing the orientations of a CME MFR from the  GCS modeling and the \insitu{} reconstruction at 1 AU. MFR rotation in the corona has also been observed during  CME eruption. \citet{green2007} find that the direction of the MFR rotation is determined by the sign of helicity of the source region. For positive (negative) helicity, the MFR rotates clockwise (counterclockwise). This implies that the conversion of twist into writhe in a kink-unstable magnetic flux rope is a possible mechanism for the rotation \citep{kliem2012}. As the orientation of MFR  is  a crucial parameter  determining its geo-effectiveness, numerical simulations have been employed to understand the physical mechanism for the MFR rotation. \citet{torok2003,torok2005} find that  vortex flows at the MFR foot points can drive a significant rotation (more than 90{\degr』) in an ideal MHD model. It suggests that the kink instability can cause the rotation. Using a similar model,  \citet{fan2004} obtain a larger rotation (almost 120\degr). Applying a breakout model in which the eruption starts with a developed MFR, \citet{lynch2009} argue that the right-handed (left-handed) MFR rotates clockwise (counterclockwise) with a constant rate  of about 30{\degr}/{R$_\sun$},  reaching a rotation angle of about 50\degr{} at a height of 3.5 {R$_\sun$}. Based on observational or numerical studies, other mechanisms  that cause MFR rotation have been suggested, such as reconnection with the ambient magnetic field \citep{cohen2010,thompson2011,lugaz2011} and alignment with the heliosphere current sheet \citep{yurchyshyn2008,yurchyshyn2009}.
Interested readers are directed to \citet{manchester2017} for a review of CME MFR rotation as well as deflection.

In the present paper, we investigate the propagation characteristics of the CME MFR from AR 12468 on 2015 December 16, using  extreme ultraviolet, magnetogram, white-light and  \insitu{} observations from \sdo, \soho, \sta, and \wind. The paper is organized as follows. In section 2, we describe the solar source signatures from AIA and HMI observations, and reconstruct  the configuration of the coronal magnetic field of the source region with  a nonlinear force-free field (NLFFF) method and examine  the magnetic energy density distribution at different heights.
In sections 3 and 4, the Graduated Cylindrical Shell (GCS) method and Grad-Shafranov (GS) model  are  used to reconstruct the 3D structure of the MFR  in the  corona and at 1 AU to analyse the propagation and the rotation of the CME MFR. This study illustrates the important role of the coronal magnetic  field around AR  in the CME MFR  rotation and propagation direction.

\section{Observations of the Source Active Region}\label{source}
\subsection{AIA and HMI Observations}\label{aia and hmi}
The source region of the eruption event is AR 12468 (S15W15). It produced a halo-like CME and a C6.6 flare on 2015 December 16. The flare started at 8:25 UT and peaked at 9:00 UT. The source region was well covered by the Atmospheric Imaging Assembly \citep[AIA;][]{lemen2012} and the Helioseismic and Magnetic Imager \citep[HMI;][]{schou2012} on board the Solar Dynamics Observatory \citep[SDO;][]{pesnell2012}. AIA images the solar atmosphere at temperatures ranging from 0.06 MK to 20 MK through 10 passbands (7 EUV and 3 UV passbands). The image has a temporal cadence of 12 s (EUV) / 24 s (UV) and spatial resolution of 1\arcsec.2. The HMI provides high time resolution magnetograms with a 45 s cadence and vector magnetic field data with a 12 min cadence. Its spatial resolution is 1\arcsec.0.

Figure \ref{f1} shows \sdo /AIA  131 {\AA} and  HMI observations of the source region before the eruption. The S-shaped hot channel marked by the red dotted curve in Figure \ref{f1}(a) is regarded as the MFR producing the CME. The S-shape of the hot channel shows that the chirality of the MFR is likely right-handed \citep[e.g.][]{canfield1999,green2014}. The contours of the photospheric magnetic field along the line of sight from a \sdo / HMI   magnetogram  are overlaid on the EUV image in Figure \ref{f1}(a). According to the contours, the MFR rooted in the strong  negative polarity sunspot in the north and the strong positive polarity sunspot in the south. The orientation of the polarity inversion line (PIL) is almost from the west  to the east.   The axial field of the MFR comes from the positive polarity sunspot to the positive polarity sunspot, so the axis of MFR  points to the east (marked by the arrow in Figure \ref{f1}(b)). The angle between the orientation of the central part of the MFR and the horizontal direction (east) is about 5\degr. The MFR poloidal field comes out of the photosphere in the south and goes into the photosphere  in the north of the PIL. For such a field configuration, the  MFR should be right-handed. The MFR was  aligned with the  PIL before its eruption, which can be seen from the AIA EUV observations. At 8:00 UT, it started to expand and rise. 
 
Figure \ref{f2} shows the EUV structures of the source region during the eruption of the  MFR. The contours of  \sdo / HMI   magnetograms are overlaid on the EUV images. The MFR from  the \sdo /AIA 131 {\AA} image (Figure \ref{f1}(a))  is indicated by the black dotted curve in Figure \ref{f2}(a). The  EUV loops overlying the MFR and  the open EUV structures surrounding the MFR are marked in the AIA 171 {\AA} images. At 8:00 UT, the MFR started to rise and then erupted in the AIA EUV observations.  Figure \ref{f2}(a)-(d) show that the EUV loops at the north of the MFR were pushed away by the CME MFR from 8:10 UT to 8:40 UT. The arrow  pointing to the northeast in Figure \ref{f2}(b) and Figure \ref{f2}(c) shows the  propagation direction of the disturbance  along the open magnetic field lines. Therefore, we can roughly estimate that the CME MFR was propagating to the northeast.                                        

\subsection{Coronal Magnetic Field Reconstruction }\label{mag field recon}
We use  an NLFFF method to understand the relation between the coronal magnetic field configuration and the propagation of the CME MFR in the low corona. The NLFFF method is proposed by \citet{wheatland2000} and extended by \citet{Wiegelmann2004} and \citet{Wiegelmann2010}. We employ the NLFFF code, which  has been optimized for application to \sdo/HMI vector magnetograms \citep{Wiegelmann2012}, to extrapolate the coronal magnetic field from the observed vector magnetograms in a Cartesian domain. This extrapolation method works well by comparing the extrapolated magnetic fields with observed EUV structures as shown in previous studies \citep[e.g.,][]{vemareddy2013,vemareddy2016,guo2010,wang2014,wang2015}. Here, we bin the data to about 1\arcsec.0 per pixel and adopt a computation domain of 512 $\times$ 256 $\times$ 160 grids.

Figure \ref{f3} diplays the comparison between the EUV images and the NLFFF extrapolation. The NLFFF extrapolation results of the   source region are shown from different viewpoints. We use different colors to distinguish different magnetic flux bundles. The NLFFF reconstruction gives an MFR structure that is consistent with  the hot channel in the EUV images before the eruption at 08:24 UT. The open flux bundles (in Figure \ref{f3}(c) and (d)) coming  from the positive polarity sunspot in the south of the MFR look like a wall inclining to the northeast. The strong magnetic pressure from the positive sunspot may change the propagation direction  of the MFR as well as its  axis orientation. 

We now calculate the magnetic energy density $ \varepsilon _\mathrm{B} = \mathbf{B}^2/2\mu_0 $, using the reconstruction results, where the coronal magnetic field $\mathbf{B}$ is from the NLFFF extrapolation and $\mu_0 $ is the vacuum permeability. Figure \ref{f4} plots the magnetic energy density distribution around the CME source region overlying the MFR before the eruption at heights of 1.07 {R$_\sun$} and 1.14 {R$_\sun$}, respectively. The black and white diamonds mark the footpoints of the MFR on the solar surface. The arrow shows the orientation and position of the central part of the MFR before the eruption that likely impact the Earth. The distribution of the magnetic energy density shows a strong gradient descent toward the northeast. Comparing the magnetogram contours in Figure \ref{f3}(a)  with the magnetic energy density distribution at 1.07 {R$_\sun$}, we can see that the peak of the energy density distribution corresponds to the sunspots in the south of the MFR. It suggests that the gradient descent is largely produced  by the  sunspots in the south of the MFR. The magnetic field surrounding the MFR may influence the propagation direction and axis orientation of the MFR. The present  magnetic energy gradient suggest that the erupting MFR would propagate toward the northeast.

\section{MFR characteristics in the extended corona}\label{mid corona}
During  2015 December 16 CME, the \sta{ } spacecraft was 166.5{\degr} east of  the Earth and at a distance of 0.96 AU from the Sun. Communications with \stb{} were lost from 2014 October 1. We use the beacon images of \sta{} as  the science data on 2015 December 16 are not available. The CME is observed as a halo-like CME by \soholasco. In order to determine the propagation direction of the CME and the orientation of the MFR in the extended corona, we  employ simultaneous two-point (\sta {} and \soholasco) white-light observations with the graduated cylindrical shell (GCS) model \citep{thernisien2006, Thernisien2009}. It has six free parameters: the  longitude and latitude of the propagation direction, the tilt angle of the flux rope, the height of the CME leading front, the half-angle between the two legs of the flux rope, and the aspect ratio  of the CME flux rope. The coronagraph observations and the modeling of the CME are displayed in Figure \ref{f5}. The wireframe rendering obtained from the GCS model  is consistent with the observed images from the two viewpoints. The resulting parameters are listed in Table \ref{tab:gcs}. The propagation direction of the CME deviates by about 45{\degr} to the east  and about 30{\degr} to the north from the radial direction of the source region (S15W15). This change is consistent with what the gradient of the coronal magnetic energy distribution predicts. The tilt angle of the MFR from the GCS model is --80{\degr}. The axis of the MFR may point to the south or north. In any case, the axis orientation of the MFR has changed in the corona by comparing  with the orientation of the MFR in the source region. These results reveal that the rotation and non-radial motion of the CME MFR occurred very close to the sun as suggested by \citet{kay2013,kay2015}. \citet{vourlidas2011} also shows a rotation during the early evolution of the associated CME.

\section{MFR Characteristics at 1 AU and Relation to Geomagnetic Activity}\label{mc}
The \insitu{} measurements associated with the CME on 2015 December 16  at \wind{} are presented in Figure \ref{f6}. A shock passed \wind{} at 15:40 UT on 2015 December 19. A magnetic cloud (MC) is  identified with the plasma and magnetic field parameters. The magnetic field is measured in RTN coordinates (in which R points from the Sun to the spacecraft, T is parallel to the solar equatorial plane and points to the planet motion direction, and N completes the right-handed triad). The  magnetic field strength first increases to about 20 nT and then decreases smoothly. The T component is largely positive, the R component rotates from positive to negative, and the N component rotates from negative  to  positive. The  D$_\mathrm{st}$ profile indicates a two-step geomagnetic storm sequence with a global minimum of --155 nT. The first dip is produced by the southward magnetic field component in the sheath region behind the shock, while the second one is caused by the southward field within the MC that is as high as --20 nT and lasts for about 24 hours. The modeled D$_\mathrm{st}$ index using the \citet{om2000} formula (minimum --156 nT) agrees with the observed  D$_\mathrm{st}$ well, where the \citet{burton1975} method gives a deeper globe minimum (--244 nT) than observed.

We use a Grad-Shafranov (GS) method \citep{hs1999,hs2002}  to reconstruct the MC structure, which has been validated by well-separated, multi-spacecraft measurements \citep{liu2008gs}. The GS reconstruction also helps determine the boundaries of the MC. There is also a significant southward field component ahead of the  resulting MC leading edge (see Figure \ref{f6}). The magnetic field strength, however, is not smooth. The GS method is sensitive to the boundaries. When interval with southward fields starting from about 3:00 UT to 16:00 UT on 2015 December 20 is included, the GS method cannot give a reasonable reconstruction. The  reconstruction result are shown in Figure \ref{f7}. The right-handed flux-rope structure resulting from the GS reconstruction is consistent with the MFR structure in the source region. It has an elevation angle of about --70{\degr} and an azimuthal angle of about 100{\degr} in RTN coordinates of the Earth.  The maximum value of the axial magnetic field is 17.7 nT, comparable to the total magnetic field strength. The axial magnetic field component is the main contributor to the southward magnetic field producing  the  geomagnetic storm, given the almost southward axis orientation.

Figure \ref{f8} shows the  axis orientations of the MFR from different observations. The axis elevation angle and  azimuthal angle of the MFR  at 1 AU are  --70{\degr} and 100{\degr}, respectively (OC in Figure \ref{f8}). The axis of the MFR  at 1 AU largely points southward. If the axial magnetic field did not change its sign during the propagation in the inner heliosphere, the  orientation of the MFR from the GCS model would also largely point to the south. The aixs orientation of the MFR at 1 AU is thus almost  aligned with the orientation of the MFR in the extended corona (OB in Figure \ref{f8}).  Comparing with the MFR orientation in the source region (OA in Figure \ref{f8}), we can see that the axis has rotated by about 95{\degr}. These results indicate that the rotation of the MFR may have occurred mainly in the low corona. The MFR axis was largely horizontal in the source region but became almost southward in the extended corona and at 1 AU. It is the rotation of the flux rope that has resulted in the intense geomagnetic storm. 

\section{summary and discussions}\label{summary}
We have investigated the rotation and non-radial motion of a CME MFR from  AR 12468 (S15W15) on 2015 December 16, using the EUV, magnetogram, white-light and \insitu{} observations from \sdo, \soho, \sta, and \wind. We use an NLFFF model to extrapolate the coronal magnetic field arroud the AR, a GCS method  to determine the propagation direction and orientation of the MFR near the Sun, and a  GS reconstruction technique  to derive the magnetic structure of the MFR at 1 AU and understand  its relationship to the geomagnetic activity. The results are summarized and discussed as follows.

1. The CME changed its propagation direction by about 45{\degr} in latitude and about 30{\degr} in longitude in the low corona due to the asymmetric distribution of the magnetic energy around  the source region. We obtain the coronal magnetic field of the AR from the  NLFFF method. The open magnetic fields surrounding the MFR  incline to the northeast of the source region. In order to evaluate the influence of the magnetic pressure on the CME propagation direction, we analyse the the magnetic energy density distributions at different heights based on the NLFFF extrapolation results. The magnetic energy density distribution on each layer shows a  gradient descent toward the northeast of the AR, which is consistent with the CME propagation direction. This indicates that the coronal magnetic field context of the AR plays an important role in the CME propagation direction.

2. The MFR  rotated counterclockwise by about 95\degr{} in the low corona during the eruption, which, again, can be attributed to  the coronal magnetic field configuration around the AR. The error in the MFR orientation in our case cannot be rigorously determined, but is roughly 20\degr{} at the source, 20\degr{} in the corona from the GCS method, and  10\degr-30\degr{} from GS reconstruction. These errors may not affect our conclusion on the MFR  rotation here, which is about 95\degr. Compared with  the MFR on the solar surface  before  eruption, the  MFR obtained from the GCS model in white-light observations rotated by about 95{\degr} counterclockwise. The near-Earth MFR from GS reconstuction has almost the same orientation  as  the MFR  from  the GCS model. The MFR was nearly horizontal in the solar source region, but became largely southward in the extended corona and near the Earth. Therefore, the rotation of the MFR plays an important role in the generation of the intense geomagnetic storm. Previous studies of the kink stability \citep{green2007} and of the magnetic tension force \citep{lynch2009} suggest that the rotation direction of an MFR is determined by the sign of helicity of the source region: the positive (right-handed) ones rotate clockwise while  the negative (left-handed) ones rotate counterclockwise. In our case, the  MFR is right-hand but rotates counterclockwise. However, the possibility of a clockwise rotation for 265{\degr}, which would result in the same orientation as in the extended corona, can not be completely excluded. In any case, the geometry of the MFR in  relation to the AR coronal magnetic field context is considered to be an important factor for the rotation.
 
\acknowledgments
The research was supported by the Recruitment Program of Global Experts of China, NSFC under grants 41374173, 41774179  and 41604146, and the Specialized Research Fund for State Key Laboratories of China. We acknowledge the use of data from {\soho, \stereo, \wind, \sdo}, and the D$_\mathrm{st}$ index from WDC in Kyoto.

\clearpage

\begin{deluxetable}{ccccccccc}
	\tablecaption{CME Parameters Obtained from the GCS Model. \label{tab:gcs}}
	\tablehead{
	  & \colhead{Time(UT)} & \colhead{Lon($^{\circ}$)} 
		& \colhead{Lat($^{\circ}$)} & 
		\colhead{Tilt Angle($^{\circ}$)} & \colhead{Aspect Ratio} & 
		\colhead{Half Angle($^{\circ}$)} & \colhead{{Height (R$_\sun$)}}
	}
	\startdata
	 & 11:12 & E30.0 & N15.0 & --80.0 & 0.4 & 35.0  & 10.0  \\
	 & 12:30 & E30.0 & N15.0 & --80.0 & 0.4 & 35.0  & 14.0  \\
 	\enddata
\end{deluxetable}
 \clearpage 
 
\begin{figure}
	\epsscale{1.2}
	\plotone{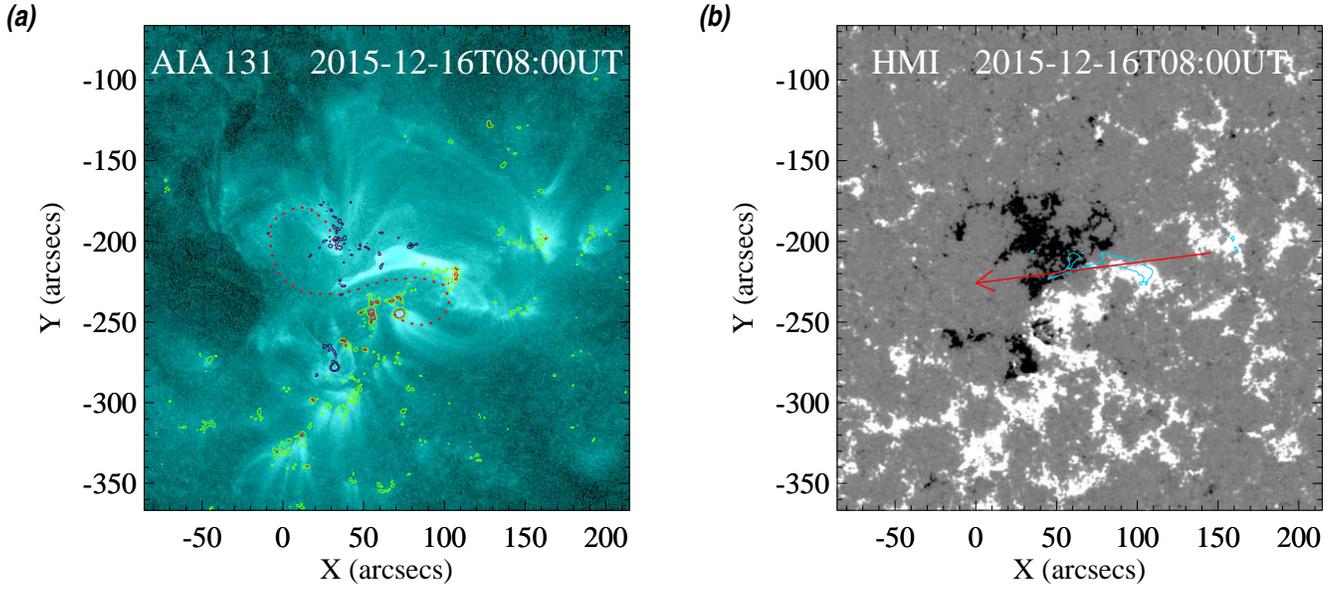}
	\caption{\label{f1}EUV and HMI observations of the source region before the eruption. (a): \sdo/AIA 131 {\AA} EUV image of source region. The dashed red curve indicates the synoptic configuration of the flux rope. Contours of --800, --400, 400 and 800 G of the photospheric magnetic field along the line of sight from a HMI magnetogram is overlaid on the EUV image and are marked in black, blue, green and red, respectively.  (b): Line-of-sight magnetogram from \sdo/HMI. The  blue contour indicates the central part of MFR. The red arrow shows the orientation of the central part of the  MFR. }
\end{figure}

\begin{figure}
	\epsscale{1.2}
	\plotone{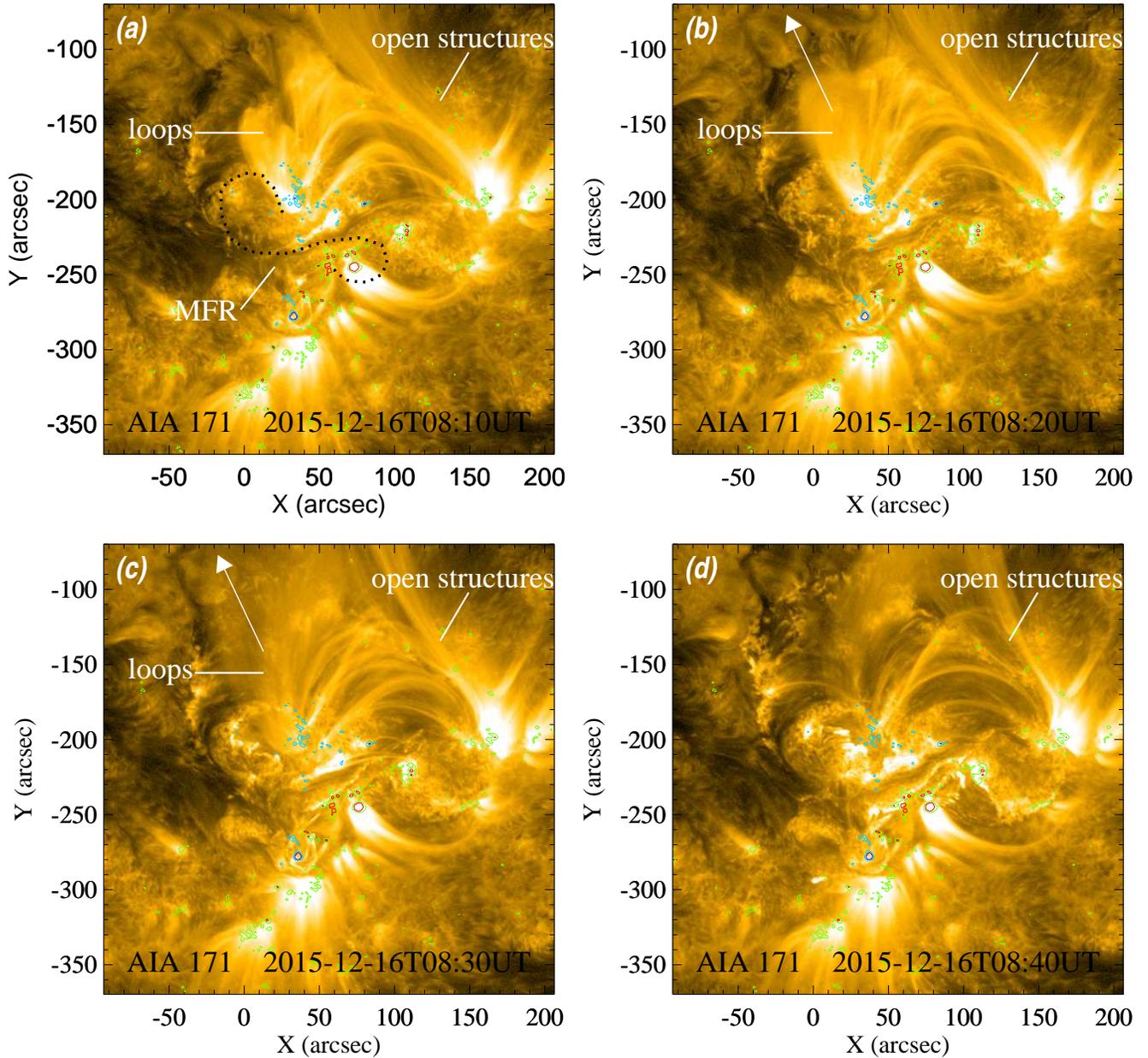}
	\caption{\label{f2}EUV structures  and the eruption of the CME MFR in \sdo/AIA 171 {\AA} at different times. Contours of --800, --400, 400 and 800 G of the photospheric magnetic field along the line of sight from  HMI magnetograms are overlaid on the EUV images and are marked in blue, cyan, green and red, respectively. The black dotted curve in (a) shows the location of the MFR from \sdo/AIA 131 {\AA}. The loops overlying the MFR and the open EUV structures surrounding the MFR  are marked respectively in (a)-(c). The arrow in (b) and (c) shows the expansion direction of the overlying loops. The open EUV structures are also marked in (d).    }
\end{figure}

\begin{figure}
	\epsscale{1.15}
	\plotone{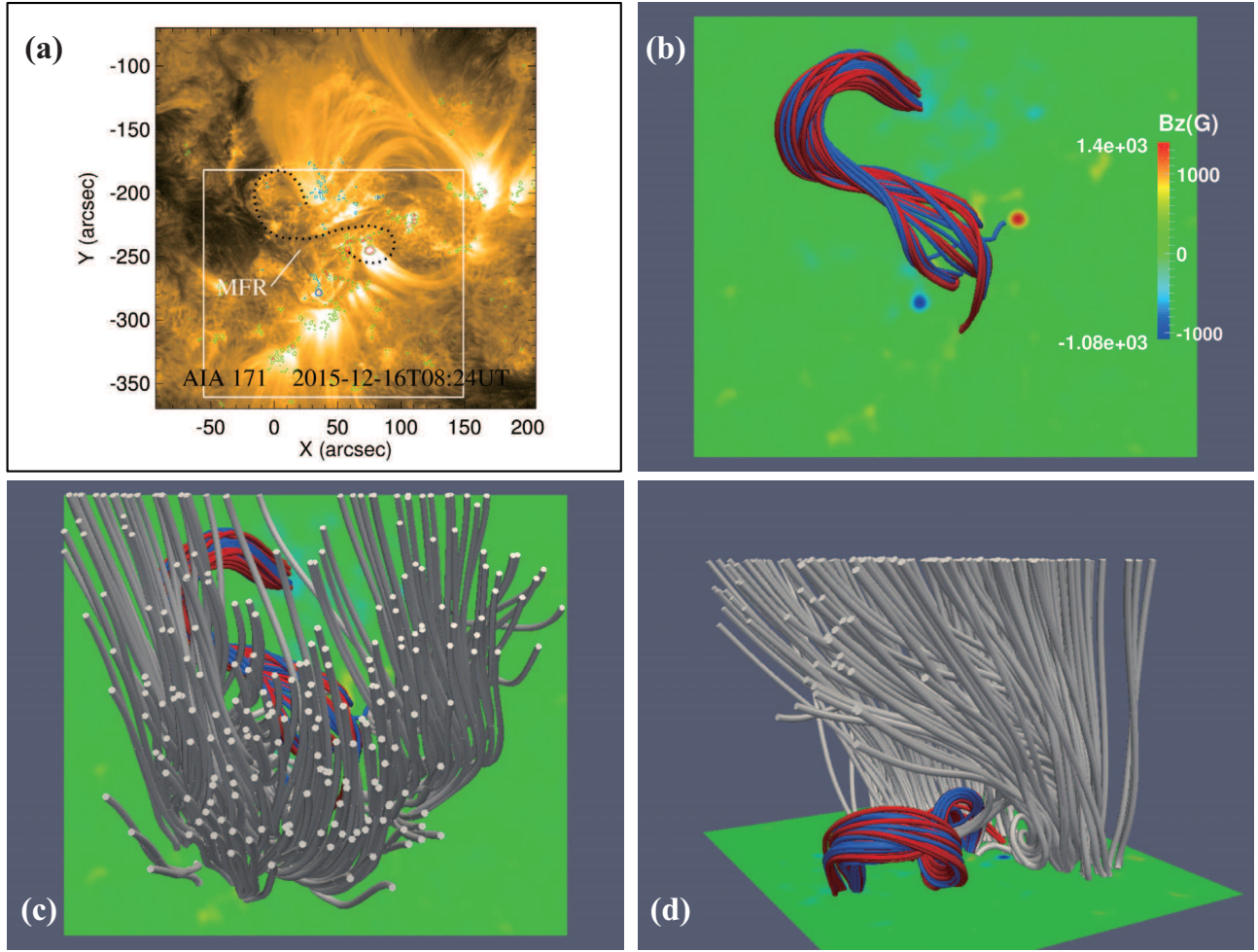}
	\caption{\label{f3}EUV structures of the source region from \sdo/AIA 171 {\AA} and the NLFFF reconstruction of coronal magnetic fields of the source region at 8:24 UT. (a): Contours of the photospheric magnetic fields are the same as in Figure \ref{f2}.  The white box indicates the reconstruction region for (b)-(d). (b) and (c): The red and blue field lines represent the MFR. The white lines show the open flux bundles. The color bar gives the vertical component of the background magnetic fields of the photosphere. (d): A view from the northeast of the AR.	
	}
\end{figure}

\begin{figure}
	\epsscale{0.8}
	\plotone{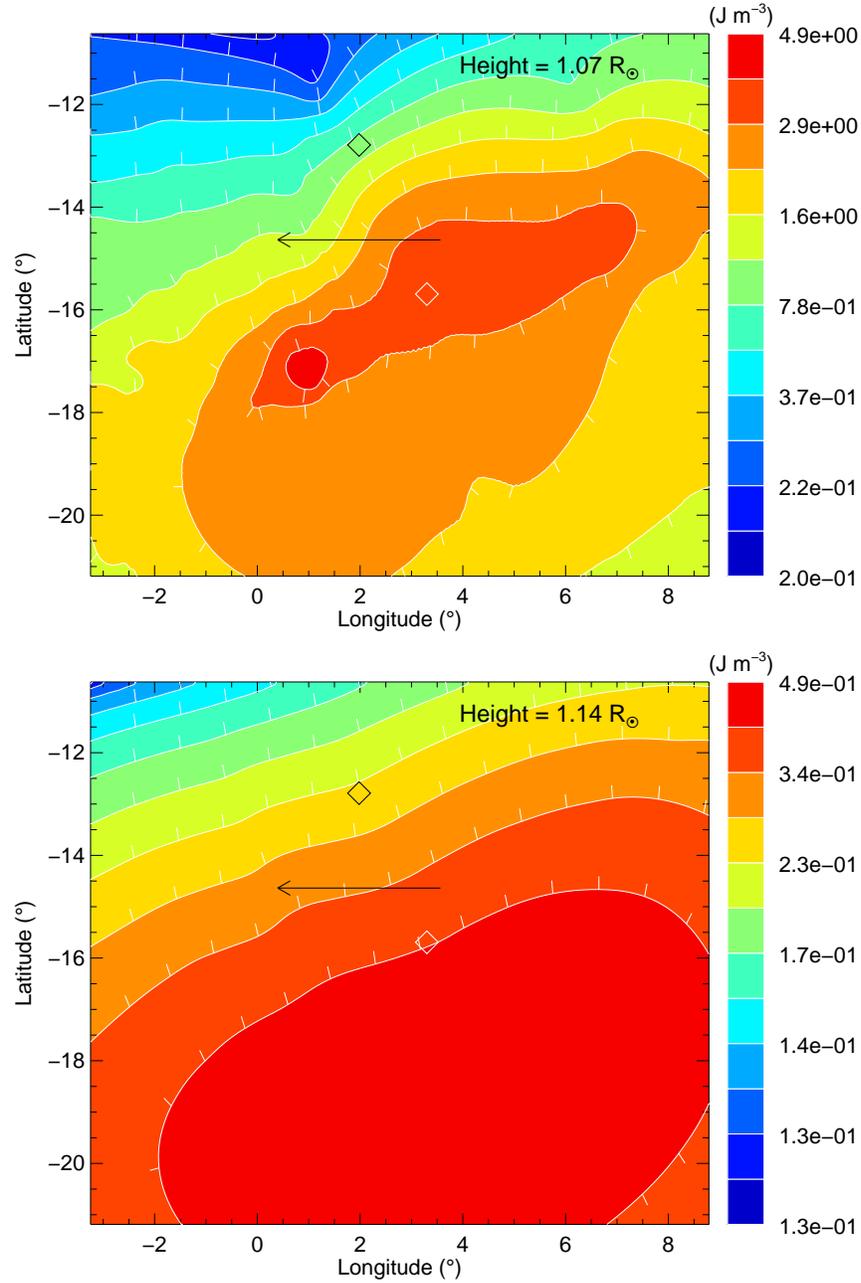}
	\caption{\label{f4}Distribution of the magnetic energy density at the heights of 1.07 {R$_\sun$} (upper panel) and 1.14 {R$_\sun$} (lower panel) based on the NLFFF reconstruction. The black and white diamonds mark the footpoints of the MFR on the solar surface. The arrow shows the orientation of the central part of the MFR before the eruption.  }
\end{figure}

\begin{figure}
	\epsscale{1.0}
    \plottwo{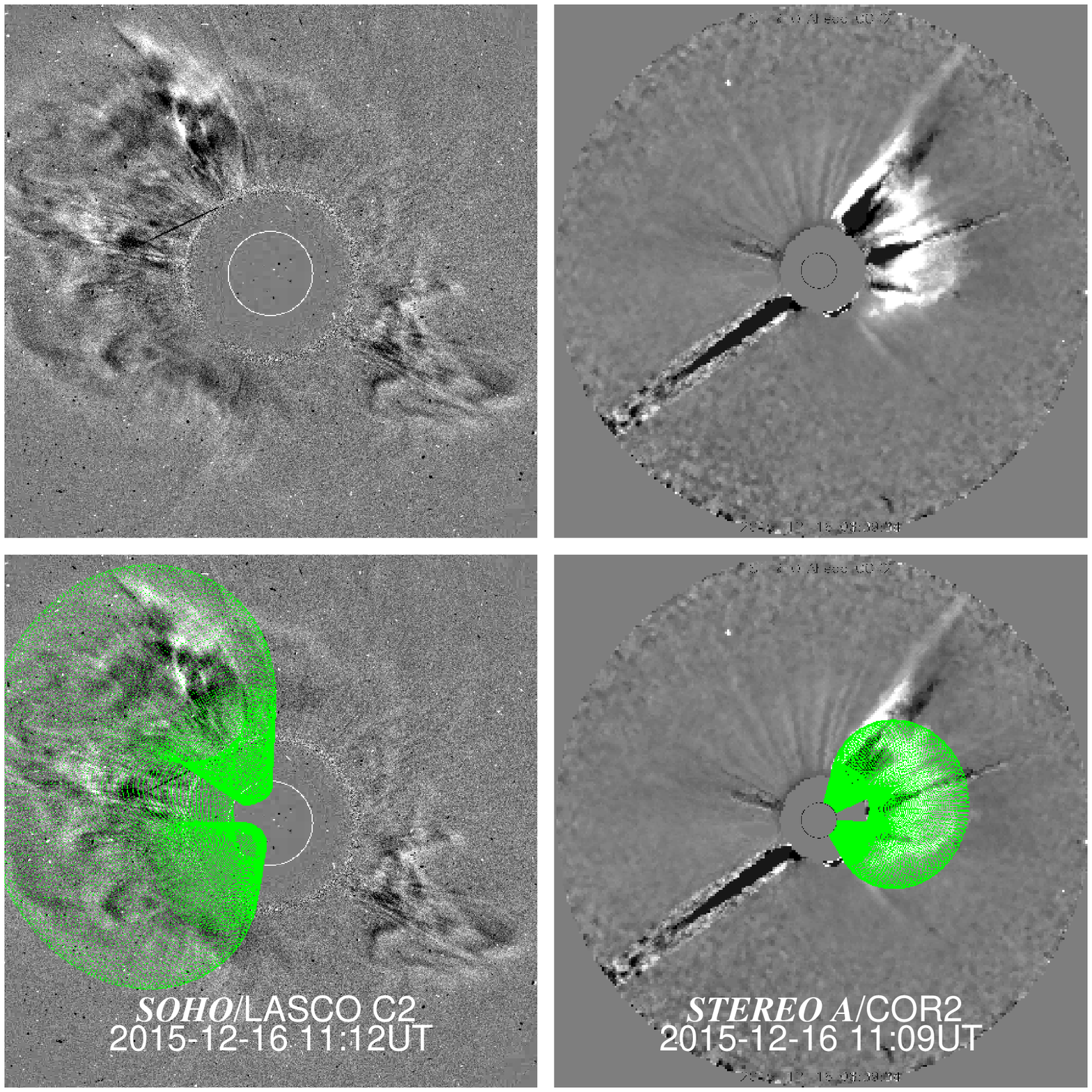}{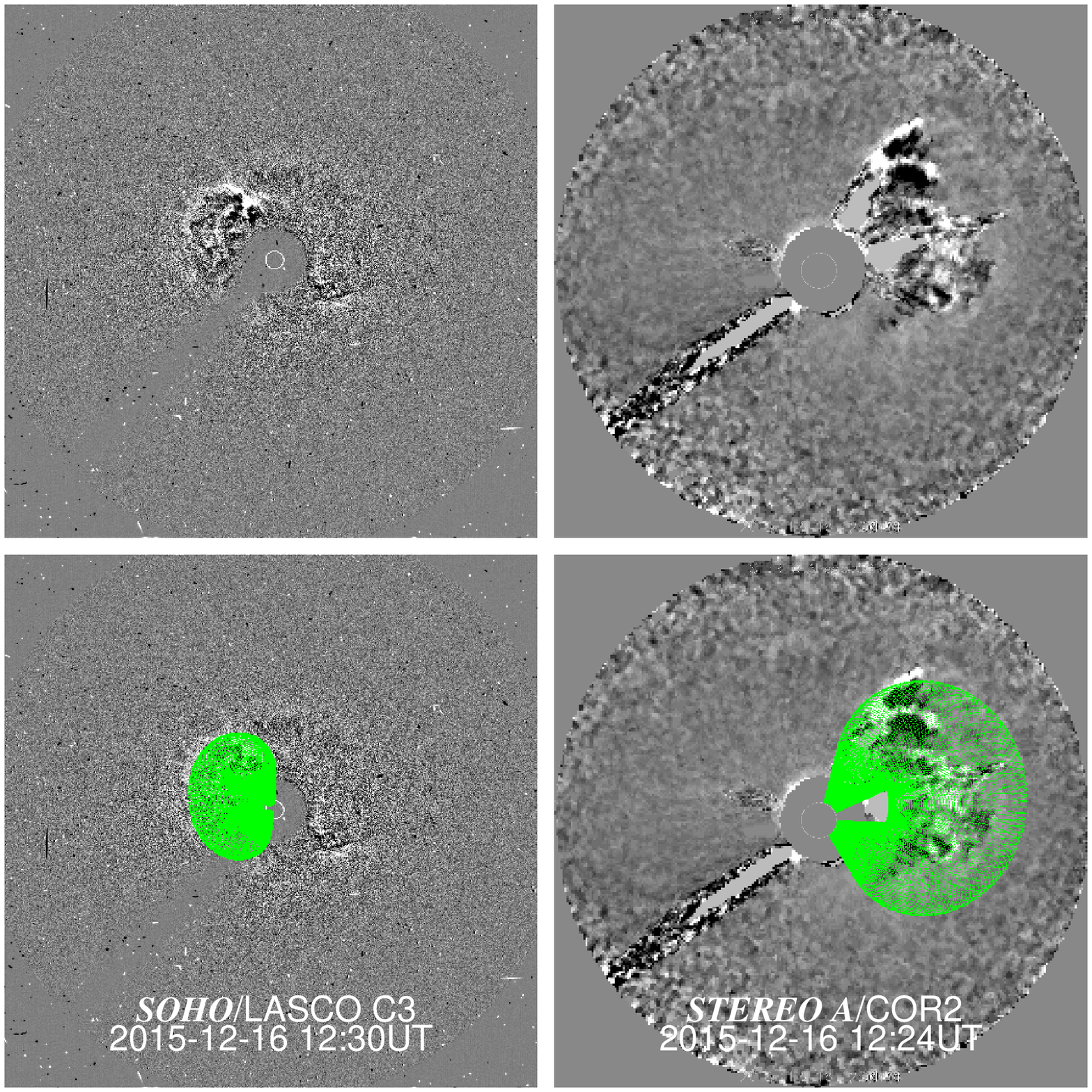}
	\caption{\label{f5}Running difference coronagraph images of the CME from \soholasco{} and \sta/COR2 (upper) and GCS modeling overlaid on the observed images (bottom). Only two frames are available for \sta/COR2 observations. }
\end{figure}

\begin{figure}
	\epsscale{0.8}
	\plotone{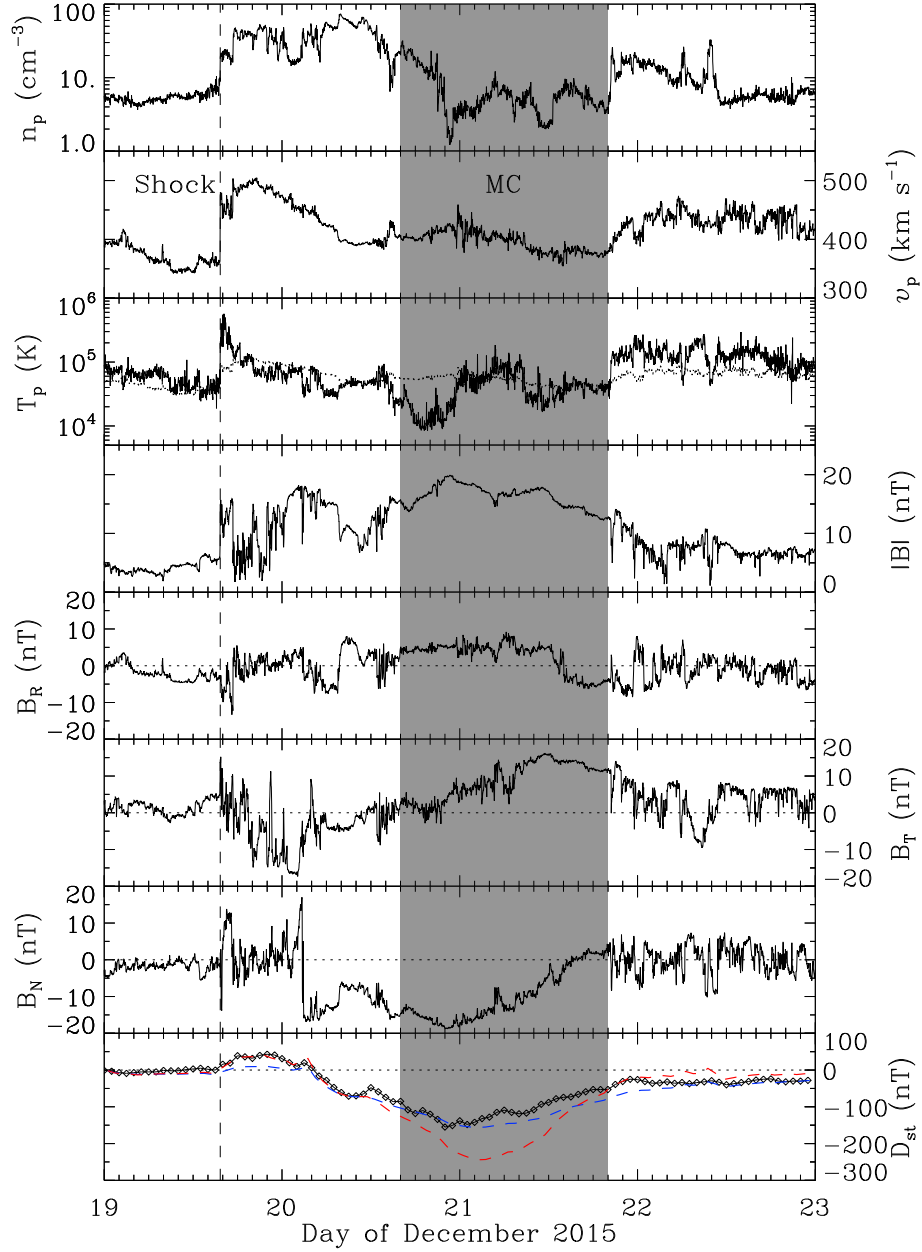}
	\caption{\label{f6}Solar wind plasma  and magnetic field parameters	associated with the CME  at \wind. From top to bottom, the panels show the proton density, bulk speed, proton temperature, magnetic field strength and components,	and D$_\mathrm{st}$ index, respectively. The dotted line in the third panel denotes the expected proton temperature calculated from the observed speed	 \citep{lopez1987JGR}. The blue and red dashed curves in the bottom panel represent D$_\mathrm{st}$ values estimated with the southward magnetic field component in GSM coordinates using the formulae of \citet{om2000}  and \citet{burton1975}, respectively. The shaded region indicates the magnetic cloud interval determined by the GS reconstruction, and the vertical dashed line marks the associated shock.}
\end{figure}

\begin{figure}
	\epsscale{0.9}
	\plotone{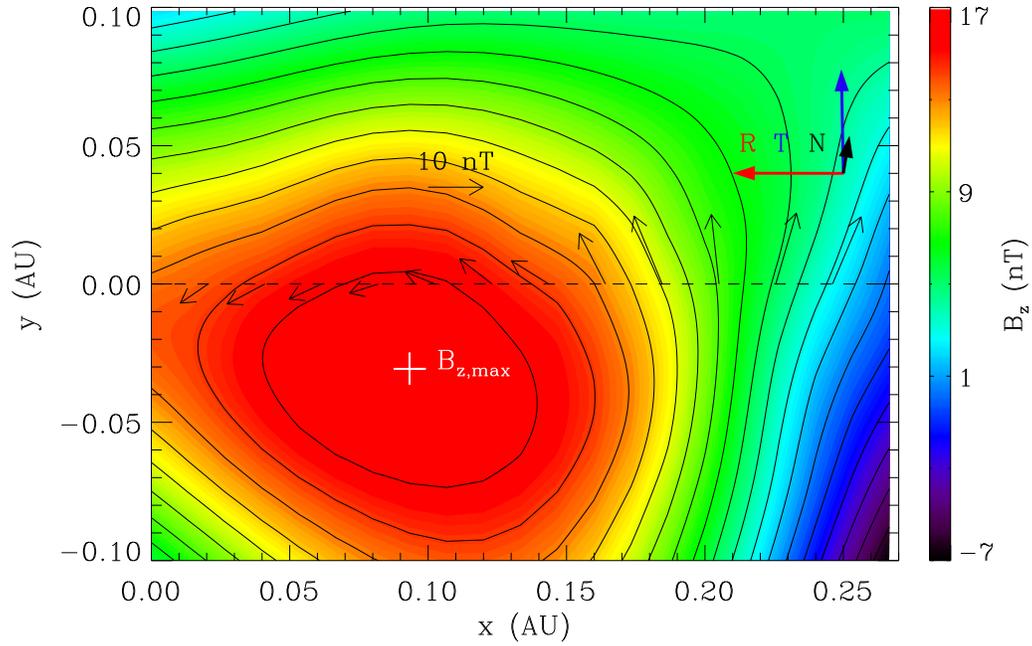}
	\caption{\label{f7}Reconstructed cross section of the magnetic cloud at \wind.
	Black contours show the distribution of the vector potential, and the color shading indicates the value of the axial magnetic field	scaled by the color bar on the right. The location of the maximum axial field is indicated by the white cross.
	The dashed line marks the trajectory of \wind. The thin black arrows denote the direction and magnitude of the observed magnetic fields projected onto the cross section, and the thick colored arrows show the projected RTN directions. }
\end{figure}
\begin{figure}
	\epsscale{0.8}
	\plotone{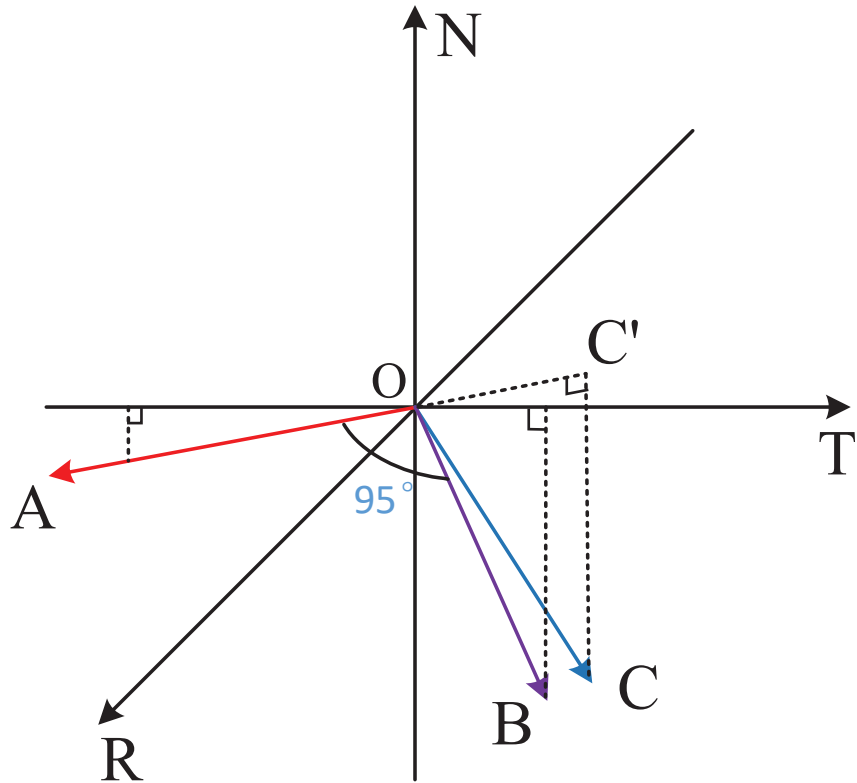}
	\caption{\label{f8}Axis orientations of the flux rope from different observations. The red (OA), purple (OB) and blue (OC) arrows show the orientations of the MFR in the source region, in the extended corona, and at  1 AU in RTN coordinates of the Earth, respectively. OA has an elevation angle of --5{\degr} and azimuthal angle of 270{\degr}. OB has an elevation angle of --80{\degr} and azimuthal angle of 90{\degr}. OC has an elevation angle of --70{\degr} and azimuthal angle of 100{\degr}.		}
\end{figure}

\clearpage
\bibliography{references}
\bibliographystyle{aasjournal}
\end{document}